\documentclass[review,sort&compress,11pt]{elsarticle}
\usepackage{hyperref,soul,multirow}
\usepackage[utf8]{inputenc}
\usepackage{amsmath,amssymb,color,xcolor,gensymb}
\usepackage{graphicx,caption,subcaption} 
\usepackage{amsfonts,amsthm,amsthm,MnSymbol} 
\usepackage{graphicx,float} 

\usepackage{xcolor}
\usepackage{geometry}
\begin{document}
\begin{frontmatter}

\title{Effect of Sn on generalized stacking fault energy surfaces in zirconium and its hydrides}

\author{P. Chakraborty$^1$}
\author{I. Mouton$^2$}
\author{B. Gault$^{1,3}$}
\author{A. Tehranchi$^1$}
\author{J. Neugebauer$^1$}
\author{T. Hickel$^{1,4}$}
\address{1-Max-Planck-Institut f\"ur Eisenforschung GmbH, D-40237 D\"usseldorf, Germany }
\address{2-Université Paris-Saclay, CEA, Service de Recherches Métallurgiques Appliquées, 91191, Gif-sur-Yvette, France}
\address{3-Department of Materials, Imperial College, South Kensington, London SW7 2AZ, UK}
\address{4-BAM Federal Institute for Materials Research and Testing, 12489 Berlin, Germany}

\begin{abstract}
Hydrogen embrittlement in Zr alloy fuel cladding is a primary safety concern for water based nuclear reactors. Here we investigated the stabilisation of planar defects within the forming hydrides by Sn, the primary alloying element of Zircaloy-4 used in the cladding. In order to explain formation of hydrides and planar defects observed in our experiments, we performed atomic-scale \textit{ab initio} calculations focusing on the solute interactions with generalized stacking faults in hcp $\alpha$-Zr and fcc zirconium hydrides. Our calculations showed that an increase in Sn concentration leads to a stabilisation of stacking faults in both $\alpha$-Zr and hydride phases. However, the solution enthalpy of Sn is lower in the $\alpha$-Zr as compared to the other hydride phases indicative of two competing processes of Sn depletion/enrichment at the Zr hydride/matrix interface. This is corroborated by experimental findings, where Sn is repelled by hydrides and is mostly found trapped at interfaces and planar defects indicative of stacking faults inside the hydride phases. Our systematic investigation enables us to understand the presence and distribution of solutes in the hydride phases, which provides a deeper insight into the microstructural evolution of such alloy's properties during its service lifetime. 
\end{abstract}


\end{frontmatter}


\section{Introduction}
Zirconium-based alloys are commonly used as fuel cladding material in the core of nuclear power reactors because of their low thermal neutron capture cross section and good corrosion resistance~\cite{Lemaignan1994}. While in operation, the fuel cladding is in contact with water, which promotes the oxidation of Zr~\cite{Proff2011}. This process releases free hydrogen, part of which enters the alloy and gives rise to the formation of hydrides once the solid solubility limit has been exceeded~\cite{ridley}. This can have a detrimental effect on the integrity and longevity of the material, as it can lead to the degradation of the mechanical strength through various defect nucleation processes and defect evolution leading to embrittlement~\cite{chapman2017}. The wide use of zirconium alloys in nuclear power reactors has motivated numerous studies on pure zirconium and zirconium hydrides over the past few decades~\cite{ridley, keevers, Won2015, Northwood}. These large efforts to elucidate hydride properties are related to the ingress of hydrogen during service in pressure and boiling water reactors, during transport and storage. At low temperatures and pressures, zirconium is known to have a hexagonal-close-packed (HCP) structure that is commonly referred to as the $\alpha$-phase. The low H solubility limit in zirconium and the thermodynamics of the Zr-H alloy system  favors the formation of brittle hydrides, which gives rise to severe safety issues~\cite{intro1,intro2,intro3,puls1986elastic,perovic1982role,bai1994hydride,altobelli2014hydrogen}. Three main hydride phases have been reported. 
The $\delta$-phase is a non-stoichiometric phase with a face-centered-cubic (FCC) structure, in which the regular lattice sites are occupied by the Zr atoms and the tetrahedral interstitial sites are randomly occupied by hydrogen. Depending on the temperature, the hydrogen content of the $\delta$-${\rm ZrH}_x$ phase varies in the interval 
1.4 $< x <$ 1.7. Increasing the H content beyond this upper limit leads to the formation of $\varepsilon$-${\rm ZrH}_2$, which is a face-centred tetragonal (FCT) structure with a unit cell having $c < a$. Another well-known phase is the $\gamma$-ZrH phase, which is a FCT structure with $c > a$, usually considered as metastable. It 
is also frequently observed in non-equilibrium precipitation
after quenching from a solid solution. Recent works~\cite{intro4,intro5,intro6} have identified a new hydride phase $\zeta$ probably
with a composition close to ${\rm {Zr}_2H}$ and metastable like $\gamma$.

Hydride nucleation and growth processes involve  dislocation behaviors, including dissociation, spreading motion, recombination and cross-slip, in the basal plane of the $\alpha$-Zr lattice~\cite{Zhang2016,ridley,carpenter1973,cross-slip2021}. Recent observations by atom probe tomography (APT) on Zircaloy-4 following electrochemical hydrogen charging motivated the extensive ab-initio calculations of the present work. We sought to explain the distribution of the primary alloying element of Zircaloy-4, Sn, in a microstructure containing  a hcp Zr lattice and with zirconium hydrides by focusing in particular on the solute interactions with stacking faults. Our results shed light onto the stabilisation of crystalline defects by alloying elements during the growth of hydrides that can further contribute to the material's embrittlement.

\section{Materials and methods}

\subsection{Experimental Details}
Details of the experiments underlying this work can be found in Refs. \cite{Ruth, isa}. In short, plates of Zircalloy 4 with a blocky-$\alpha$ microstructure \cite{Viv} were electrochemically charged by using a solution of 1.5 wt.~\% H$_2$SO$_4$ in H$_2$O at 65 $^\circ$C for 24 h (all details are in \cite{Ruth}). The thick hydride surface layer was dissolved and the hydrogen homogenised throughout the alloy by annealing at 400 $^\circ$C for 5 h. The sample was left to cool in the furnace cooling at approximately 0.5 $^\circ$C/min to promote the formation of both intra- and intergranular hydrides. Specimens for atom probe tomography analysis were prepared using a FEI Helios dual-beam xenon plasma focused ion beam (PFIB), using the protocol proposed in Ref. \cite{Thompson}, with the final stage of sharpening performed at cryogenic temperature to avoid hydride formation during the preparation as reported previously \cite{Yanhong, isa}. APT analyses were performed on a CAMECA LEAP 5000 XR, operated in laser pulsing mode, with specimens maintained at a base temperature of 50 K. 

\subsection{Computational Details}
The theoretical investigations were designed to be compared directly to the experimental results. The calculations were carried out using projector augmented wave (PAW) potentials as implemented in the Vienna Ab initio Simulation Package (VASP)~\cite{vasp1,vasp2}. The PBE~\cite{pbe} exchange-correlation functional was chosen. A plane-wave cutoff of 500 eV is considered for all calculations. The convergence tolerance of atomic forces is 0.01 eV/~{\AA} and of total energies it is 10$^{-6}$ eV. The \textbf{k}-point sampling was conducted by a $\Gamma$-centred Monkhorst-Pack scheme. The sampling density was set large enough that the convergence of total energies was within 2 meV per atom. The Brillouin-zone integration was made using the Methfessel–Paxton smearing.  
 
In the present work, bulk properties of $\alpha$-Zr such as the lattice constant and the solution enthalpy of Sn were calculated as a function of its concentrations. The solution enthalpy $\Delta E_{\rm s}$ of $m$ substitutional Sn atoms was calculated as 
\begin{flalign}
\Delta E_{\rm s}&= E({\rm Zr}_{n-m} {\rm Sn}_{m} {\rm H}_{nx})  
 -{(n-m) } E({\rm ZrH}_{x}) - m x E({\rm H})- m E({\rm Sn})
\label{eq:1} 
\end{flalign}
where $E({\rm Zr}_{ n-m} {\rm Sn}_{ m} {\rm H}_{nx})$ is the total energy of a $\alpha$-Zr or Zr-hydride system consisting of $n-m$ Zr atoms and $m$ substitutional Sn atoms. The chemical potential of hydrogen, $E$(H), has been determined as the concentration dependent energy difference in the $\delta$-hydride. The same chemical potential of hydrogen has been assumed for the other hydrides, because its determination is more complex in ordered phases. Alternatively, the hydride reference could be replaced by $-n E({\rm ZrH}_{x}) + m E ({\rm Zr})$, shifting the reference to the chemical potential of Zr. Eventhough, changing the two references changes the absolute values for the solution enthalpy by up to 1 eV, we have ensured that our qualitative results do not depend on the specific choice. 
The energy $E({\rm Sn})$ is determined for the $I4_{1}$/amd $\beta$-Sn system, but this choice is not relevant when comparing the Sn solublity in different phases. Full volume relaxation was performed in all calculations.

A slab supercell was constructed, separated by 10~{\AA} of vacuum from its periodic image, to calculate the generalized stacking fault (GSF) energies. The slab 
is divided into two blocks of equal atomic layers. When the upper block is subject to a relative displacement by an arbitrary vector, $\mathbf{f}$, parallel to the slip plane as shown in Fig.~\ref{fig:fig01}(a), the shifted structure will have a fault energy  $E(\mathbf{f})$. When various slip displacements, $\mathbf{f}$, are applied in the plane of the cut, $E(\mathbf{f})$ generates a surface which represents the fault-energy map against the applied displacements. This fault-energy map is the GSF energy surface
\begin{flalign}
E_{GSF}({\mathbf{f}})&=\frac{E({\mathbf{f}})-E({\rm {\mathbf{f = 0}}})}{A}    
\label{eq:2} 
\end{flalign}
where $A$ is the area of the glide plane of the supercell. The size and shape of the supercell were never changed with slip displacements. 
For further reference, the coverage ratio $C$ is defined for  substitutional atoms adjacent to the slip plane. For example, when $C$ is equal to 0.25 in Fig.~\ref{fig:fig01}a then one out of four atoms in the glide plane is substituted.

Three hydrides are studied in this work, the metastable $\zeta$ phase, which later transforms into $\delta$-${\rm ZrH}_x$, is also observed in experiments~\cite{hydride}. The $\delta$-phase is known to have an FCC structure with tetrahedral interstitial sites being randomly occupied by hydrogen. To model such compounds at finite temperatures, one can employ configurational cluster expansion techniques~\cite{lit-nat,lit-52}. The special quasi-random structure (SQS) method is sufficient when the focus is on the fully random alloy, i.e., if the impact of short-range order can be assumed negligible. The properties predicted by the SQS approach will be closer to the properties of the true random alloy with increased SQS size. Therefore, for simplicity and consistency with the other considered phases in this work, we approximate the $\delta$-phase to be a disordered alloy modelled based on ${\rm ZrH}_2$ hydride, where the H atoms occupy only six tetrahedral sites keeping the remaining two tetrahedral sites vacant. In other words, the hydride phases ZrH$_{x}$ are studied as a function of the H content $x$. 

\section{Results and Discussions}

\vspace{3mm}

\subsection{Experimental Results}

\begin{figure}
	\centering
\includegraphics[scale=.8]{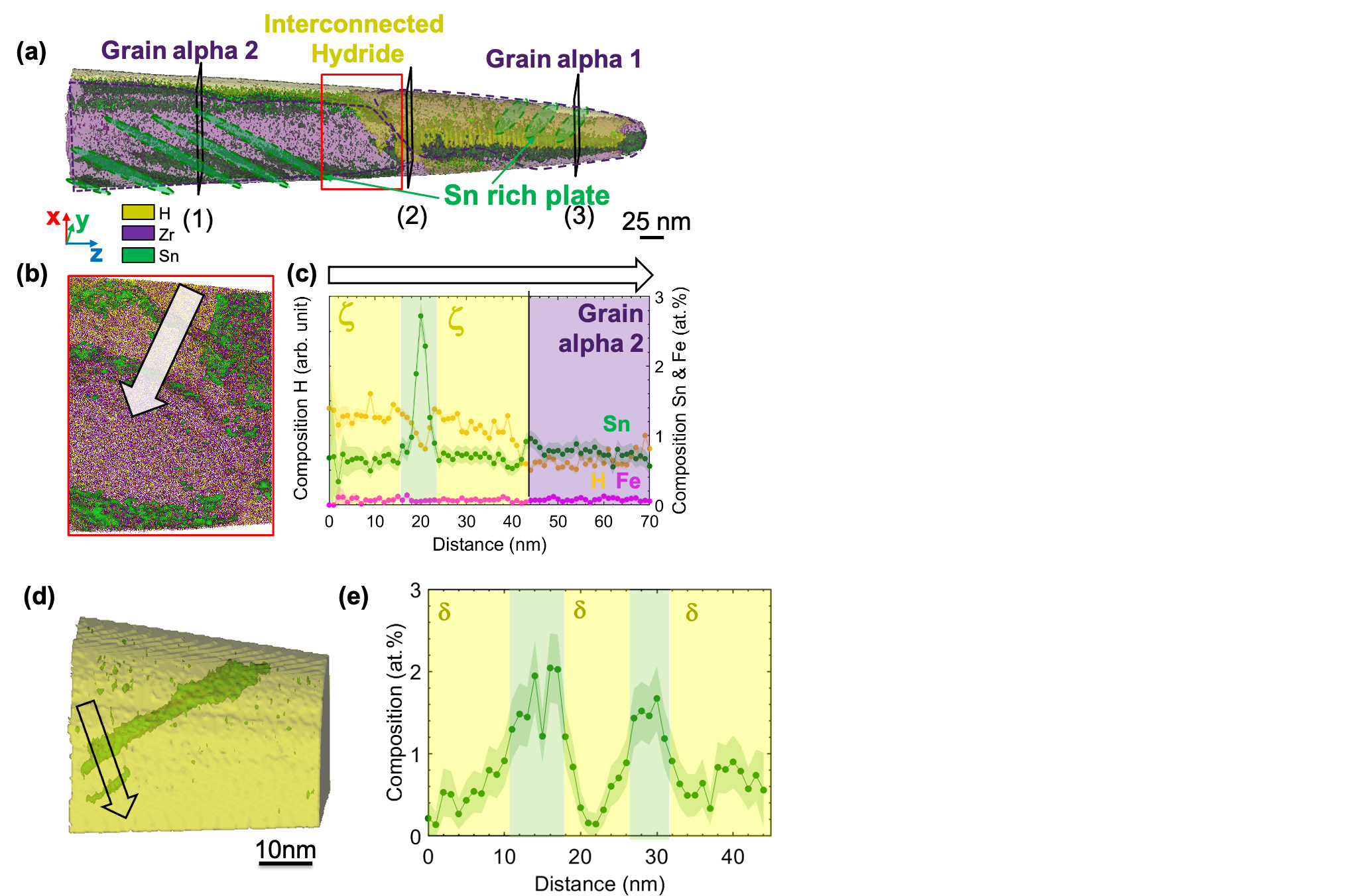}%
\caption{a) APT reconstructed image of a sample that shows a hydride pocket in between the two grains denoted as alpha 1 and alpha 2. b) the zoomed portion of the APT needle where the composition profile of Sn is evaluated. c) Composition profile of Sn taken perpendicularly to the Sn-rich plates which shows the Sn rich regions in the $\zeta$ hydride. d) another region of the APT needle is shown which contains the trapped Sn rich plates (shown in green) inside the interconnected hydride regions(shown in yellow). e) the corresponding composition profile of Sn is shown which is trapped inside the $\delta$ hydride.}\label{fig:APT}
\end{figure}

Figure~\ref{fig:APT} shows an APT dataset from the same set of analyses as reported in Ref.~\cite{isa}. This data was obtained from the region ahead of the growth front of an intergranular hydride that has grown perpendicularly to a GB. Sn is repelled from the growing hydrides and segregates in a few layers of the blocky-$\alpha$ Zircaloy-4 matrix at the interface to the $\delta$-hydrides. This agrees with existing experimental studies~\cite{Andrew, isa} that also reported the repulsion of Sn from hydrides and the presence of an intermediate $\zeta$ Zr-hydride layer at the interface between the $\delta$-hydride and the $\alpha-$Zr matrix. 
We have observed a series of planar features at the hydride/matrix interface, which appear segregated with Sn and were interpreted as stacking faults in the $\alpha-$Zr matrix~\cite{isa}, as readily visible in Fig~\ref{fig:APT}a. 

Hence, the composition profile in Fig.~\ref{fig:APT}c shows that while Sn is repelled locally from the hydrides as they grow, it eventually gets trapped in  planar features present in the hydrides. Fig.~\ref{fig:APT}e demonstrates that the trapping of Sn to planar features is not limited to $\zeta$ hydride, but applies to $\delta$ hydride as well.  It is further documented  from transmission electron microscopy (TEM) that the stacking faults start growing ahead of the hydrides. The crucial question we ask here is the reason behind such an increase in accumulation of Sn in planar features, while the solubility inside the hydrides is apparently small (Fig.~\ref{fig:APT}). Henceforth, the interaction of Sn with the $\alpha$-Zr matrix and with the hydrides are evaluated separately in presence of stacking faults and the comparison is used to explain the mechanisms behind the observations.

\subsection{Simulation Results}

\subsubsection{${\alpha}$-Zr and Sn solid solution} 

To understand and explain why our experiments show a lower solubility of Sn inside the hydrides as compared to the Zr matrix, but at the same time a substantial increase in the Sn concentration along planar features we have performed ab initio simulations. 
As a first step, the bulk phase of ${\alpha}$-Zr and Sn solid solution are considered. 
Table~\ref{tab:bulk} shows the lattice constants of $\alpha$-Zr and $\beta$-Sn obtained using the present DFT calculations, which are in good agreement with reported values in the literature~\cite{carpenter, lit-2}. All supercells have been constructed by using these lattice parameters as initial values. Subsequently, the effect of changing the atomic concentration of Sn from 1.04 to 8.33 at.\% has been studied.
The solution enthalpy of Sn in ${\alpha}$-Zr is calculated to be -1.15 eV, the energy gained by the Sn atom with reference to its bulk phase. Here, the supercell is chosen such that one Zr atom is substituted by Sn in a (0001) plane containing eight Zr atoms.

In order to understand the  interaction of Sn with planar defects, we follow the above mentioned assumptions and analyze the generalized stacking fault (GSF) energies obtained by inducing a translational slip along the [01$\bar{1}$0] direction. Doing this, the system encounters a high-symmetry configuration at translation $\mathbf{b}$/3; $\mathbf{b}$ being the Burgers vector in the [01$\bar{1}$0] direction. The corresponding local minimum is commonly known as the stable stacking fault (SSF). The energy barrier that precedes the SSF is termed as unstable stacking fault (USF) energy. The SSF is of significant importance in determining the quantitative character of the plastic deformation. A reduction in the SSF indicates enhanced dissociation of the dislocation core in the basal plane, while for the USF energy it indicates a reduction of lattice resistance to dislocation motion.

\begin{table}[t]
\centering
\caption{Lattice parameters (\AA) of $\alpha$-Zr and $\beta$-Sn compared 
 with other DFT works~\cite{carpenter} and experimental values~\cite{lit-2}}\label{tab:bulk}
\begin{tabular}{c c c c c c}
\hline
& System & Present & Theory & Exp \\
\hline

&  $\alpha$-Zr (a/c) & 3.23/5.17 & 3.23/5.17 & 3.23/5.14 \\
&  $\beta$-Sn (a/c) & 5.95/3.21 & 5.94/3.21 & 5.83/3.18  \\
\hline

\end{tabular}

\end{table}

\begin{figure}
    \centering
    \begin{subfigure}[b]{0.22\textwidth}
   \centering
    \includegraphics[width=1.1\textwidth]{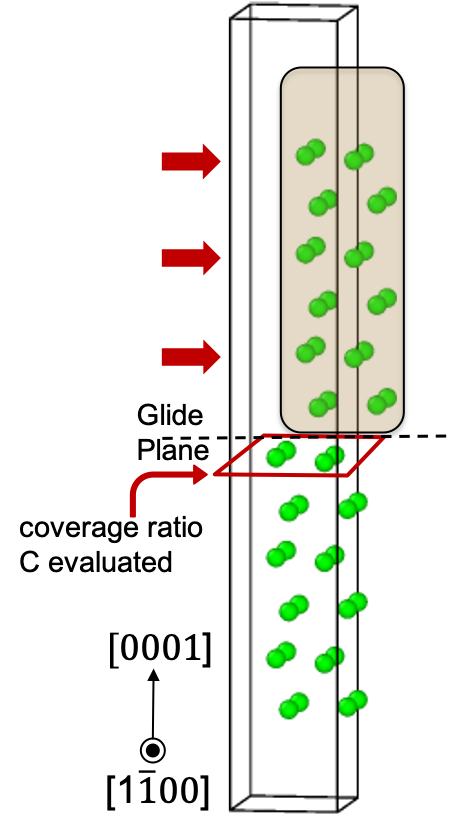}
    \end{subfigure}
\begin{subfigure}[b]{0.6\textwidth}
\centering
\includegraphics[width=1.1\textwidth]{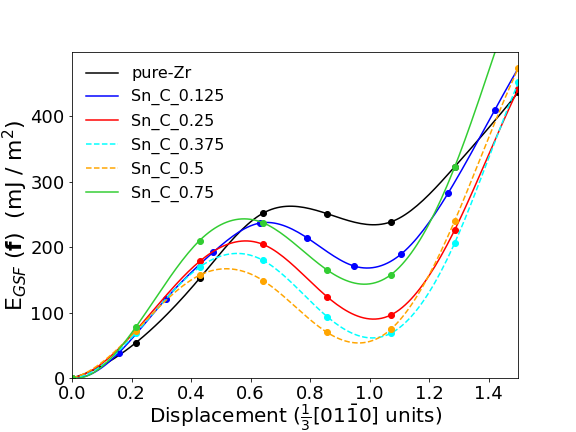}
\end{subfigure}
  \caption{(left) Schematic view of GSF calculations for the basal plane in $\alpha$-Zr. The figure represents the shifted structure when the upper part of the initial configuration is glided along the [01$\bar{1}$0] direction in the $\{$0001$\}$ basal plane. The
  coverage ratio of the glide plane, C, is evaluated inside the marked plane. (right) Calculated GSF energies along the [01$\bar{1}$0] $\gamma$ line for pure $\alpha$-Zr and $\alpha$-Zr with Sn atoms in the glide plane for different coverage ratios C as marked in the figure. The dashed lines for Sn coverage C = 0.375 and C =0.5 at the glide plane denotes the Boltzmann average at each slip displacement at 400$^{\circ}$C~\cite{Ruth, isa}. The data points are connected using spline fitting. }
  \label{fig:fig01}
\end{figure}

\begin{figure}
     \centering
     \begin{subfigure}[b]{0.46\textwidth}
         \centering
         \includegraphics[width=1.1\textwidth]{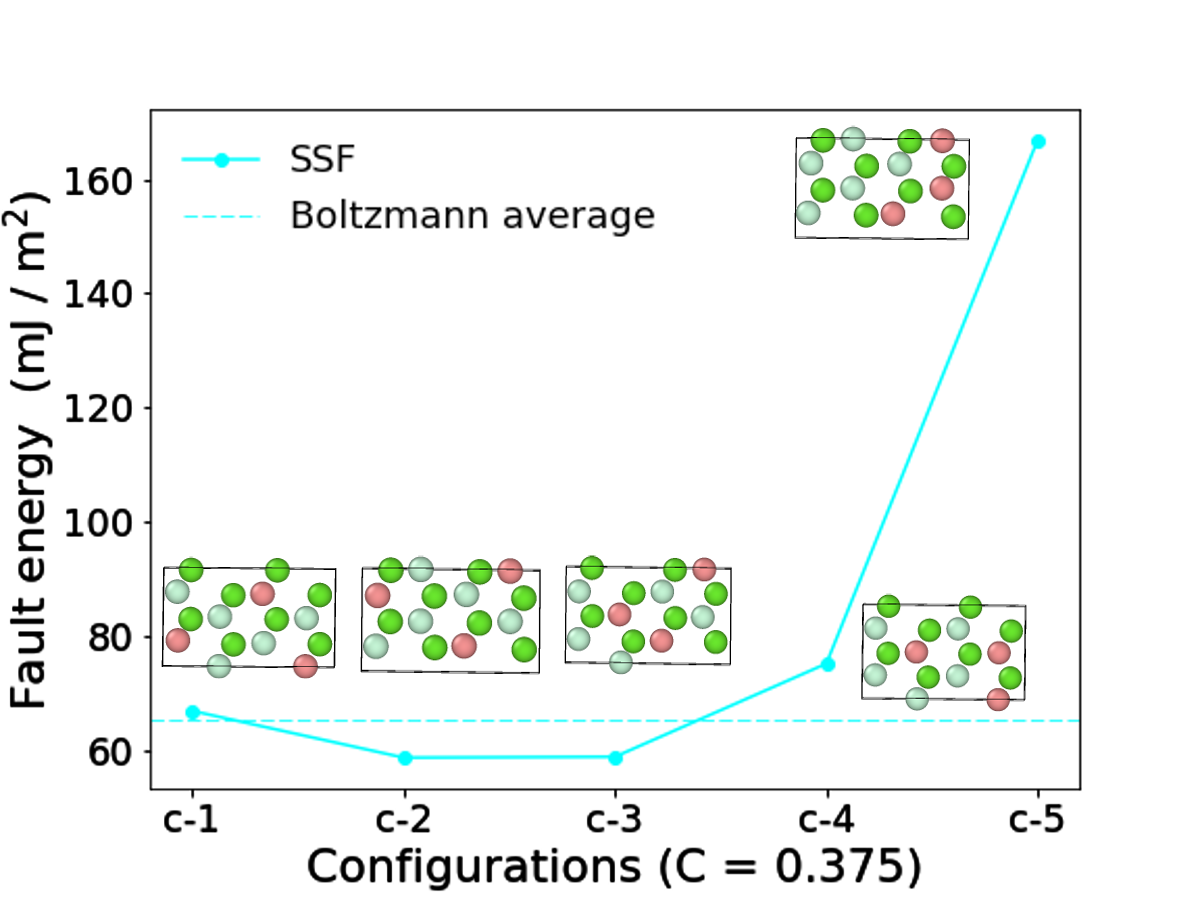}
     \end{subfigure}
     \hfill
     \begin{subfigure}[b]{0.46\textwidth}
         \centering
         \includegraphics[width=1.1\textwidth]{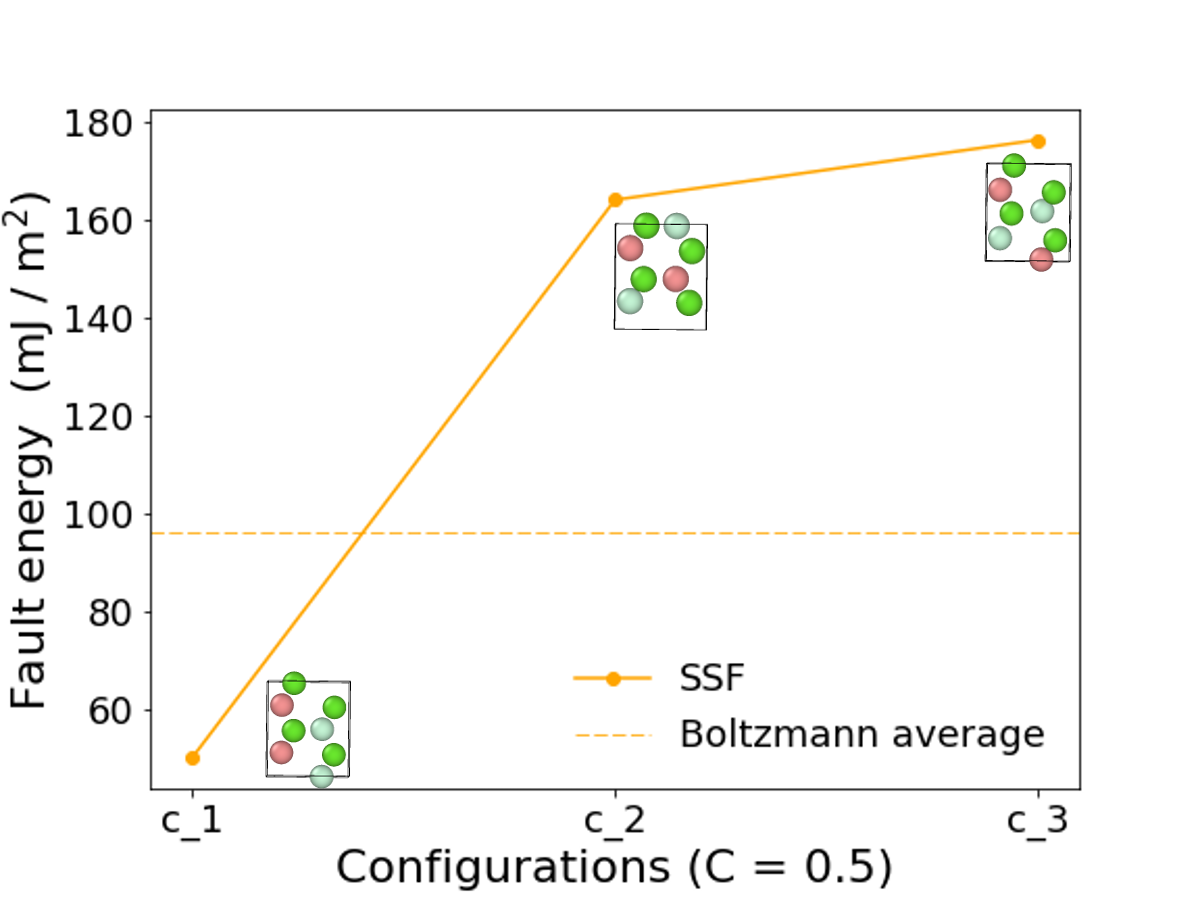}
     \end{subfigure}
        \caption{SSF energy for various configurations of Sn atoms belonging to a coverage-ratio of (left) C = 0.375 and (right) C = 0.5. A snapshot of the fault plane at the stacking fault is shown for each configuration. The green spheres represent the host Zr atoms and red spheres are the Sn atoms, two different atomic planes are shown with lighter shade corresponding to the fixed atomic plane with Sn atoms. The plot provides the stability of the SSFE of Zr–Sn binary systems with respect to a clustering of Sn atoms in the glide plane. The horizontal dashed line is the Boltzmann average of the datapoints at 400$^{\circ}$C (annealing temperature~\cite{Ruth, isa}).}
        \label{fig:fig03}
\end{figure}

The GSF properties in pure $\alpha$-Zr are in very good agreement with existing literature values~\cite{lit-4}. The impact of Sn atoms that are substituted in the glide plane only is shown in Fig~\ref{fig:fig01}. For 50\% coverage of Sn atoms at the glide plane, we have considered the results for three stable configurations as shown in Fig.~\ref{fig:fig03}. For 37.5\% coverage, a larger simulation cell is considered that is 3 out of 8 atoms on the plane is a Sn atom. Accordingly, there can be several configurations, here we have compared five configurations with different energies. For these two cases, the SFE plots in Fig~\ref{fig:fig01} represents the mean of the possible configurations for each displacement. The probability for a possible configuration is determined by the statistics of Sn segregation and is therefore approximated by a Boltzmann average at 400$^{\circ}$C, annealing temperature~\cite{Ruth, isa}
at each displacement corresponding to a particular coverage value $C$.


\begin{figure*}[!b]
	\centering
\includegraphics[scale=.55]{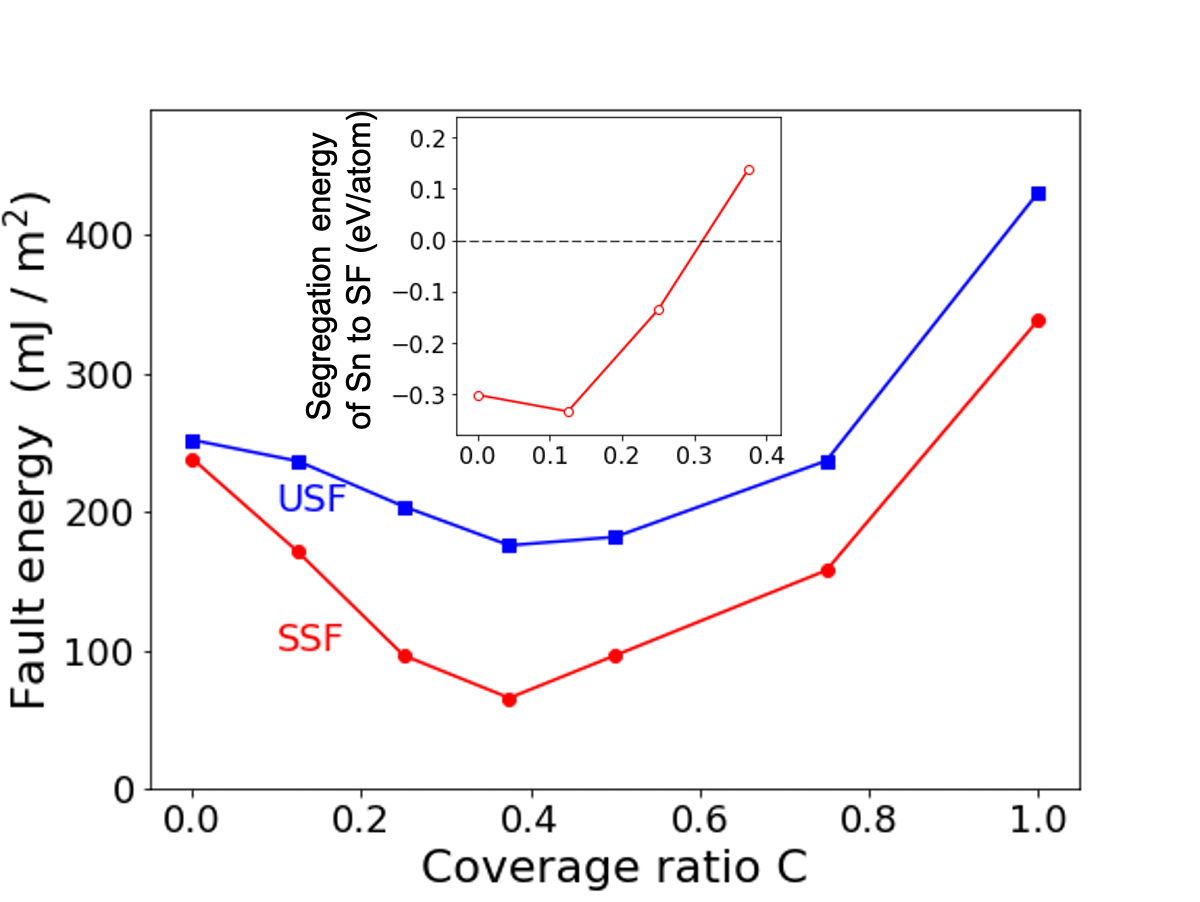}%
\caption{Fault energies of the stable stacking fault (SSF, red circles) and unstable stacking-fault (USF, blue squares) of an $\alpha$-Zr  lattice in the basal plane as a function of the coverage-ratio $C$ with Sn. The inset shows the segregation energy of Sn atoms with the corresponding coverage ratio C to the SSF, where negative values indicate attraction.}\label{fig:fig02}
\end{figure*}


The dependence of the unstable and stable stacking fault energies on Sn coverage are depicted in Fig.~\ref{fig:fig02}. With the substitution of Sn atoms in pure Zr we notice a monotonous decrease of the energy for  stacking fault configurations with $C < 0.4$. This indicates that the presence of Sn stabilizes the stacking fault structure in $\alpha$-Zr. At the same time, this means that Sn tends to segregate to these planar defects. Indeed, the solution enthalpy of single Sn in the (0001) plane with 8 atoms ($C = 0.125$) is lower at the SF (-1.45 eV) as compared to pure Zr without fault (-1.15 eV). Fig.~\ref{fig:fig01} indicates that this segregation effect is strongest for the translation $\mathbf{b}$/3 and, for example, much smaller at the USF (see Fig.~\ref{fig:fig02}). Therefore, Sn indeed stabilizes the SSF. At any of the other defect configurations along the glide plane the stabilisation effect is small and can be neglected. This is further supported by the segregation energy profile of Sn atoms with planar coverage C in the GP as seen in the inset of Fig.~\ref{fig:fig02}.
Furthermore, both USE and SFE increase again significantly towards $C = 1$, and there is an optimum coverage of the glide plane for less than half of a monolayer of Sn atoms.

\begin{figure}
     \centering
     \includegraphics[scale=.68]{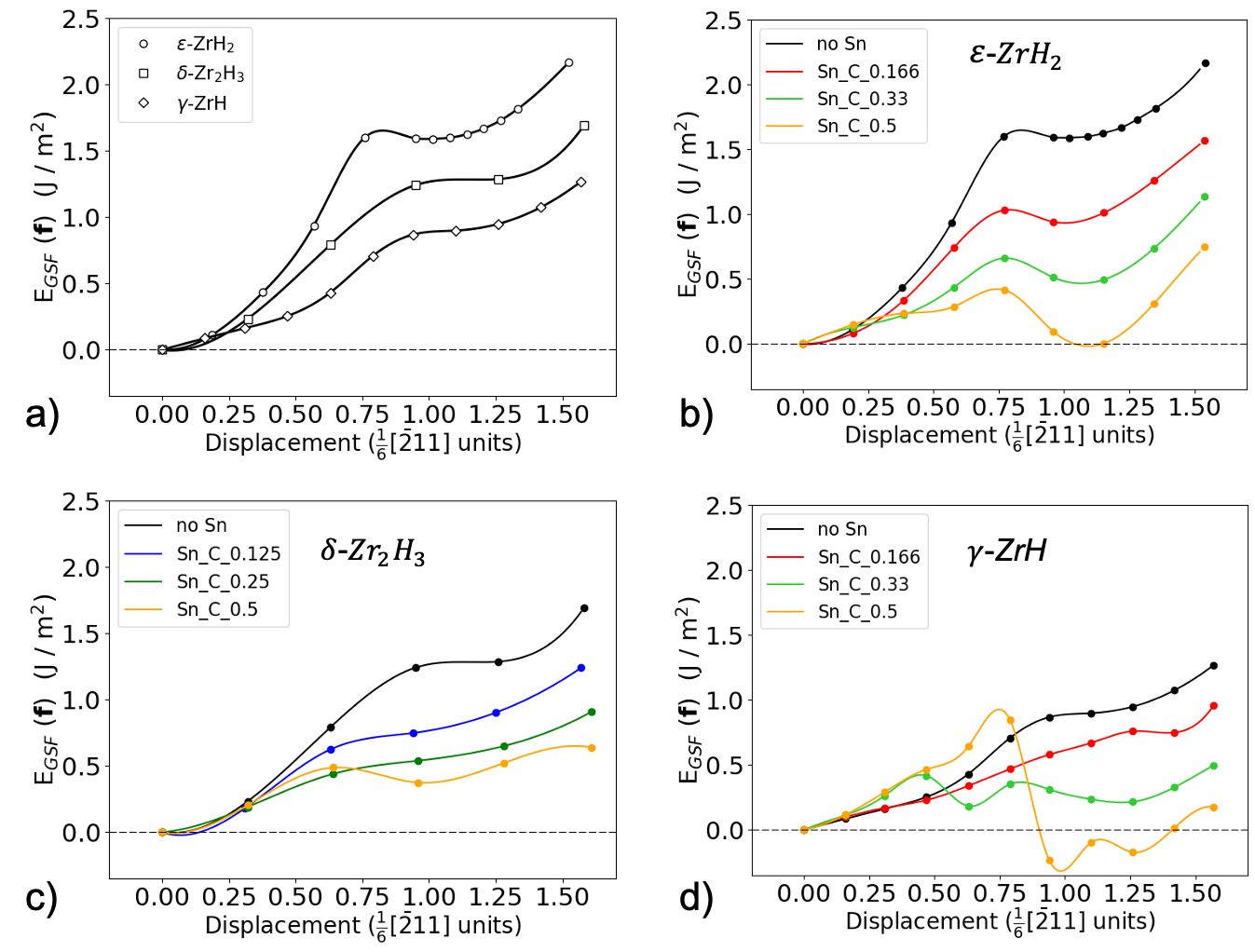}
        \caption{a)--d) GSF  energy profiles along [$\bar{2}11$] direction in the fcc $\{111\}$ plane : a)comparison of the USF energy and the SSF energy is shown in this plot between the three hydrides without any solute atoms. b) - d) GSF energy profile for $\varepsilon$-ZrH$_{2}$, $\delta$-Zr$_{2}$H$_{3}$  and $\gamma$-ZrH for pure hydride and with substituted Sn atoms with different coverage ratio in the GP are plotted respectively.}
        \label{fig:fig0a}
\end{figure}

\subsubsection{Zirconium hydrides and Sn}

\begin{figure}
     \centering
     \begin{subfigure}[b]{0.45\textwidth}
         \centering
         \includegraphics[width=1.1\textwidth]{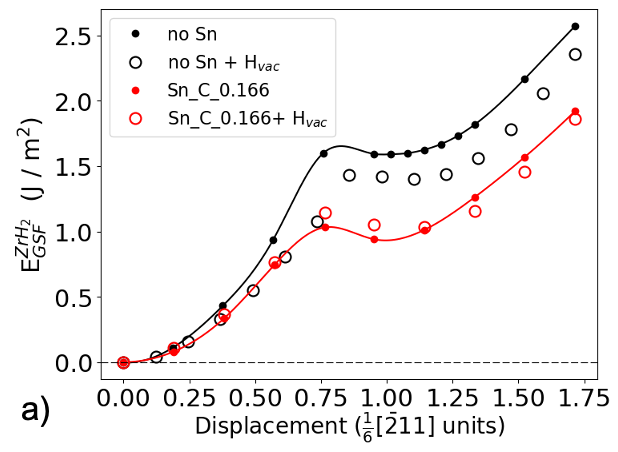}
     \end{subfigure}
     \hfill
     \begin{subfigure}[b]{0.45\textwidth}
         \centering
        \includegraphics[width=1.1\textwidth]{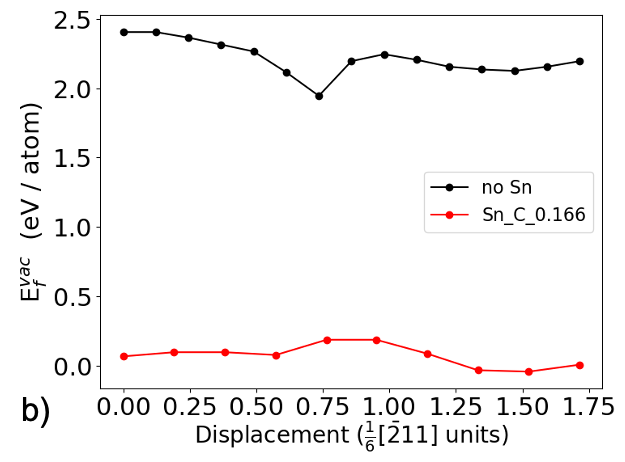}
    \end{subfigure}
        \caption{a) Influence of H vacancy in pure $\varepsilon$-ZrH$_{2}$ and hydride with Sn coverage of 0.166 at the glide plane is compared here. b) The corresponding vacancy formation energy is shown. }
\label{fig:vac}
\end{figure}

We will now focus on the effect of Sn on the GSF in the different hydride systems. Similar to the previous analysis in $\alpha$-Zr,  Zr is substituted by Sn on the GP in the hydrides for various coverage ratios. 
We first note that the calculated solution enthalpy of Sn in $\delta$-hydride is 1.81 eV, which is positive, clearly supporting our earlier statement of the larger solubility in $\alpha$-Zr, where the value was -1.15 eV. 

The GSF energy profiles are plotted along the 
$\langle \bar{2}11 \rangle$ direction in the fcc $\{111\}$ plane for the hydrides. As shown in  Fig~\ref{fig:fig0a}b, the GSF energy is found to be the highest for the $\varepsilon$-hydride followed by $\delta$-${\rm ZrH}_{1.5}$. 
This correlates almost linearly with the H concentration in these phases, while there is at the same time a linear correlation with the c/a ratio in the different phases. 
Atoms are closely packed in the fcc structure. However, H occupies all its tetrahedral positions in ${\rm ZrH}_2$, whereas $\delta$-${\rm ZrH}_{1.5}$ has in our calculations two H vacancies in the supercell. This increases the chance of a distortion of the lattice along the low energy $\{111\}$ path by reducing the repulsive forces between the H atoms observed in the $\delta$-hydride phase, rendering a lower fault energy. 

The GSF energy profiles of the three hydrides are plotted in Fig.~\ref{fig:fig0a} for glides along the fcc $(111)$ plane with and without Sn atoms. In the case of the $\varepsilon$-hydride it is seen from Fig.~\ref{fig:fig0a}b  that the GSF has the highest energies among all the other hydrides without any Sn atom, but also shows for Sn coverages in the range $C$ = 0 - 0.33 the strongest decrease of the SFE with increasing Sn concentration.  

For (unrealistically) high coverage ratios of more than 50$\%$  in the GP the lattice becomes even unstable. For the case of  $\gamma$-ZrH as shown in Fig.~\ref{fig:fig0a}d the USF and SSF are less prominent as compared to $\varepsilon$-${\rm ZrH}_2$. The introduction of Sn in the lowest concentration further results in flattening the energy curve. This may be attributed to the fact that H is only occupying 50$\%$ of the available interstitial sites thereby resulting in reduced repulsion and enhanced mobility in the lattice structure. The GSF energy profiles of $\gamma$-hydride show a less systematic dependence on the Sn content as compared to the other hydrides, since Sn increases the GSFE for small displacements and the typical decrease is only observed for the SSF as well as larger displacements. Again, an unrealistic coverage ratio of $C= 0.5$ yields a structural instability.

Since the effect of Sn is most substantial for the $\varepsilon$-${\rm ZrH}_2$ hydride and its SSF, Sn has a strong segregation effect to stacking faults in particular in regions with high H concentrations.
Also in the other hydrides with different H content, the addition of Sn yields a substantial softening of the energy profiles particularly for the SSF, but the effect is less pronounced. To convince ourselves that the amount of hydrogen is decisive for the differences between the hydrides and not, for example, the c/a ratio, we have analyzed the Sn impact in $\varepsilon$-${\rm ZrH}_2$ with and without a hydrogen vacancy. 
Fig.~\ref{fig:vac}a compares the GSF energy profiles with 1.66 atm\% coverage of Sn with the pristine case of the hydride without Sn and with/without a H vacancy at the GP. The fault energy for the pristine hydride is more strongly lowered by Sn in the absence of a H vacancy as compared to the presence of a vacancy. 

Another way of interpreting these results is the  explicit evaluation of the H vacancy formation energy of $\varepsilon$-${\rm ZrH}_2$ with and without Sn as shown in Fig~\ref{fig:vac}b. It is seen that independent of the deformation, the vacancy formation energy of H is much lower in the presence of Sn as compared to pristine $\varepsilon$-${\rm ZrH}_2$. This indicates that Sn is strongly attracted by these vacancies. The segregation to the stacking faults is, therefore, mainly a consequence of avoiding H-rich environments as present in the bulk hydride phases. 
Indeed the positive solution enthalpy of Sn in $\delta$-hydride (1.81 eV) is substantially reduced to -0.59 eV, when stacking faults are present inside the hydride.

\section{Discussions}

\begin{figure}
	\centering
\includegraphics[scale=.58]{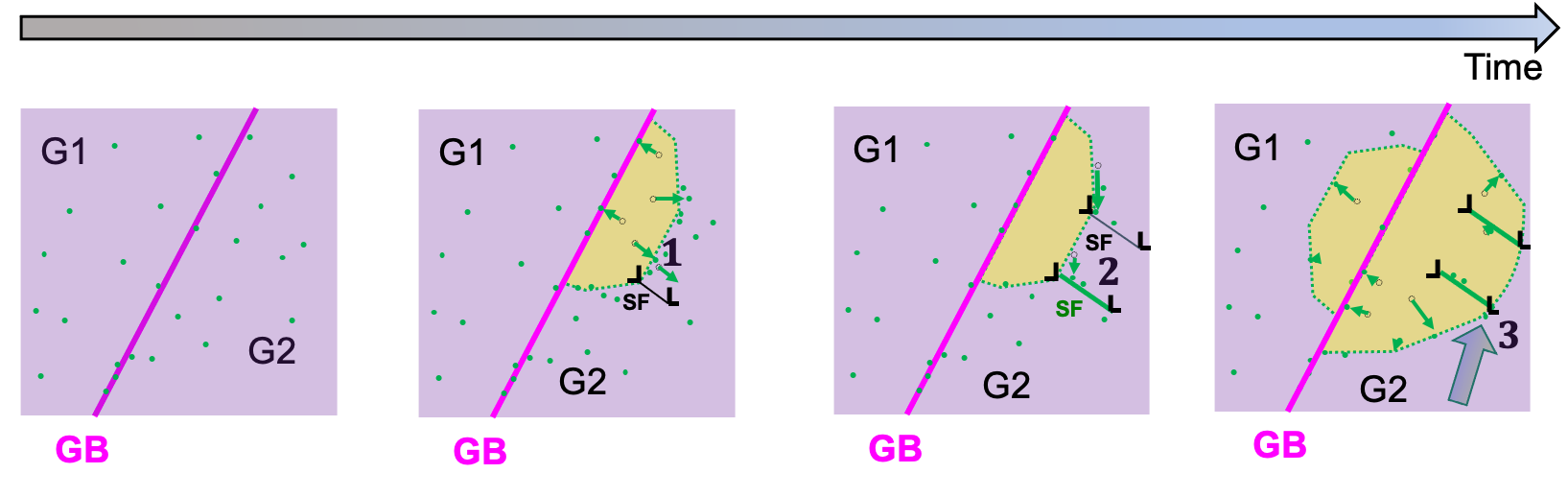}%
\caption{a)-d)Schematic illustration of a mechanism for the evolving microstructure showing a time evolution of hydrides (yellow shading) and the role of stacking faults (black symbols). The distribution and interaction of Sn is depicted by green dots and green arrows. The labels 1, 2, 3 correspond to the steps indicated in Fig.~\ref{fig:solubility}.}\label{fig:hydride1}
\end{figure}

The results outlined in the previous section suggest a complex interplay of the composition-dependent formation energies of planar defects and the segregation behaviour of Sn solutes in all relevant phases. 
Together, these insights explain the experimentally observed phenomena reported above, in particular the accumulation of Sn at planar features in the hydrides. As outlined in the last section, the direct segregation of Sn atoms solved in the hydride to the forming stacking faults, is one possible mechanism. 
The moderate annealing at 400 $^\circ$C for 5 h will, however, limit the kinetics of segregation.  

\begin{figure}
	\centering
  \includegraphics[scale=.55]{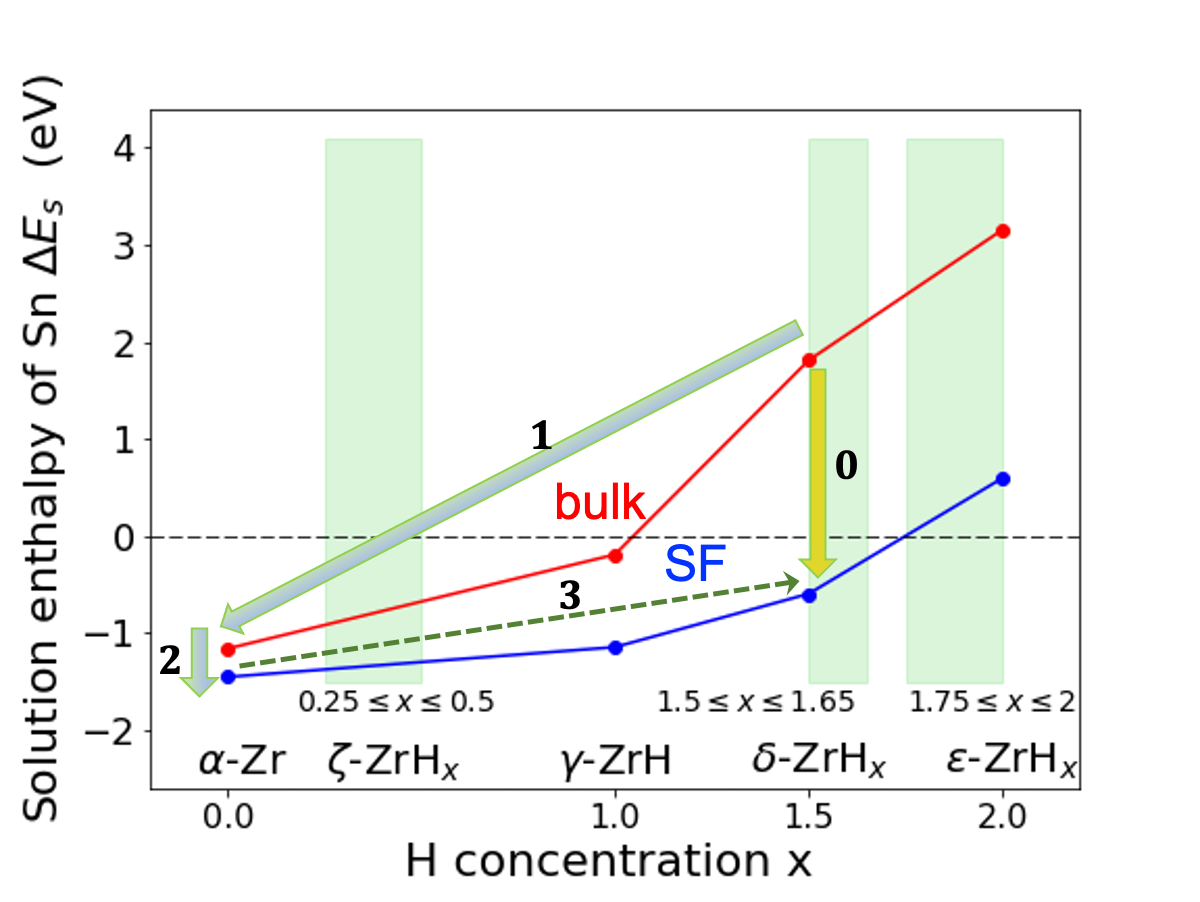}%
\caption{Solution enthalpy of Sn in pure Zr and the considered hydride phases with different H content ZrH$_x$. The solution enthalpy in the corresponding bulk phases (red symbols) and at the stacking faults (blue symbols) are compared. Two possible mechanisms are indicated by a yellow arrow with label 0 and green arrows for the steps 1, 2, and 3 underlying the schematic process in Fig.~\ref{fig:hydride1}.}\label{fig:solubility}
\end{figure}

An alternative mechanism for the microstructural evolution next to a  grain boundary (GB) is schematically depicted in Fig.~\ref{fig:hydride1}. Since hydrogen accumulates at the GB, it leads to the growth of a local hydride phase. Based on the variation in the H concentration, the experiments indicate the presence of an intermediate, metastable $\zeta$-phase in addition to the stable hydride phases. Our simulations show, however, that  the qualitative conclusions are largely independent of the hydrogen concentration. 
We consistently observe for all hydride phases that the solution enthalpy of Sn is larger than in the $\alpha$-Zr phase (Fig.~\ref{fig:solubility}). 
The resulting local Sn depletion in the region of the forming hydride can be accommodated by a Sn enrichment in the moving interface to the matrix as depicted by the green arrows in Fig.~\ref{fig:hydride1}c. This is also confirmed by the APT composition profile of Sn shown in Fig.~\ref{fig:APT}c.
It is assumed that this process is kinetically more likely than the diffusion in the bulk hydride phase itself. 

Next to the hydrides the formation of planar defects has been observed. 
The presence of stacking faults could be explained by a strain accumulation, originating from the lattice mismatch between the forming hydride and $\alpha$-Zr matrix. Wang et al.~\cite{wang} previously showed that $\alpha$-Zr grains within such hydride pockets exhibit a difference in orientation compared to the parent grain. Therefore, further growth of the hydride can be attributed to a large strain induced deformation, which causes stacking faults to propagate from the interface into the matrix. 

As the stacking faults grow into $\alpha$-Zr, they get stabilised by the presence of Sn as suggested by the above calculations. 
As a consequence, the Sn atoms that are accumulated locally at the growth front of the hydride are attracted to the stacking faults as depicted schematically in Fig~\ref{fig:hydride1}b (step 2). The DFT results have shown (Fig.~\ref{fig:fig02}) that the trapping of Sn occurs up to an optimum coverage well below 50 atm\%.  
While the hydride continues growing, this enrichment of Sn in certain layers will be transferred into this phase. 
Since an attractive interaction of Sn and SFs is  observed  for the hydride phases, the planar distribution of Sn atoms will impose the formation of stacking faults (step 3). 

The energetics of this mechanism is depicted in terms of the Sn solution enthalpy in Fig.~\ref{fig:solubility}. 
The yellow arrow with label ``0'' indicates the direct segregation of Sn atoms solved in the hydrides to stacking faults within the same phases. In the alternative multi-step mechanism, step 1 indicates the thermodynamic driving force for Sn redistribution into $\alpha$-Zr during the structural transformation from Zr to the hydride phase.  The segregation to stacking faults at the interface happens in step 2. Hence, there is an expected accumulation of Sn atoms at the interface due to the atomic rearrangements. Step 3 is the transfer of Sn from a stacking fault in $\alpha$-Zr into a stacking fault in the $\delta$-hydride. It is denoted by a dashed arrow, because the increase in solution enthalpy  indicates a missing driving force for Sn diffusion in this direction. Instead the step is determined  by the transformation of the matrix phase into the hydride phase and the subsequent formation of stacking faults therein due to the Sn enrichment. The driving force for Sn to find its way out of the hydride again (opposite direction of arrow 3), is much reduced as compared to the one in step 1. Additionally, the corresponding  kinetics of the Sn atoms is assumed to be slower as well.

The overall change in the solution enthalpy is the same for both mechanisms. However, the comparison of the solution enthalpies of Sn in stacking faults of $\alpha-$Zr and hydrides (Fig.~\ref{fig:solubility}) reveals that the driving force for Sn to stacking faults is much stronger in the hydrides (step 0) than in the Zr matrix (step 2). As a consequence, Sn atoms effectively push the SFs into the hydrides. This explains the experimentally observed planar defects covered with Sn inside the hydrides, as schematically depicted in Fig~\ref{fig:hydride1}d.  
The orientation relationship between the matrix and the hydride as observed in the experiments is
$\{0001\}_{\alpha}\|\{111\}_{\delta}$, similar to the SF planes considered for DFT calculations. 
However, since the stacking sequence of basal planes in the SSF of hcp Zr is fcc like and the stacking sequence of (111) planes in the SSF of fct hydrides is hcp like, there is not a simple continuation of the SFs between the two phases. It is more likely that the Sn enrichment stabilizes some of the hcp Zr structure in the vicinity of the original stacking faults.

In general, the introduction of Sn in hydrides lowers the SFE depending on the structural symmetry and therefore reduces the structural stability. The reduction effect by Sn is increasing with the hydrogen content in the hydride phases. At the same time, the stacking fault energy under Sn free conditions is also increasing with the Sn content. Therefore, the overall impact of Sn containing SFs on structural stability and deformation mechanisms is comparable in all hydrides.

\section{Conclusion}

In summary, using ab initio simulations of the effect of Sn on the stacking faults in hcp $\alpha$-Zr and in fcc zirconium hydrides allowed us to propose an interesting multi-step mechanisms for the microstructure formation during hydride growth in Zircaloy-4. 
The DFT simulations of the hcp $\alpha$-Zr phase showed that an increase in Sn concentration leads to a stabilisation of the stacking fault structure up to a optimal coverage of 40 atm \% of the glide plane. 
A similar attraction to stacking faults is also observed in the fcc hydride phases. 

The dominant contribution to the thermodynamic driving force of Sn to segregate to stacking faults is its interaction with H. Sn shows a much lower solubility in Zr-hydrides than in pure $\alpha$-Zr, while the presence of H vacancies improves the solubility $\varepsilon$-ZrH$_{2}$. 
The trend continues with an increasing number of H vacancies in the hydrides, yielding lower stacking fault energies in $\gamma$-ZrH  as compared to $\varepsilon$-ZrH$_{2}$. 

From the experiment it is seen that a series of hydrides is formed at each side of the GB. Small pockets of pure Zr are found along side the hydride phases. With our analysis we understand not only a Sn decoration of stacking faults, but also the reason for the formation of stacking faults inside the hydride. The chemical stabilization of the defects by Sn, which would otherwise not be present, belongs to the category of defect phases \cite{korte2022defect} that are currently reported for various materials systems \cite{vslapakova2020atomic}.

\section{Acknowledgement}

IM and BG are grateful for the Max-Planck Society and the BMBF for the funding of the Laplace and the UGSLIT projects respectively, for both instrumentation and personnel. BG and PC are grateful for financial support from the ERC-CoG-SHINE-771602. Prof. Ben Britton and Dr. Siyang Wang are gratefully acknowledged for the collaborative experimental work on Zr-hydrides and deuterides, which were prepared and analysed in depth as part of HexMat (EP/K034332/1) and MIDAS (EPSRC EP/SO1720X) programme grants. AT, TH and JN acknowledge financial support by the Deutsche Forschungsgemeinschaft within the CRC1394 ``Structural and chemical atomic complexity – from defect phase diagrams to material properties''.


\newpage

\end{document}